\RequirePackage{fix-cm}
\documentclass[twocolumn,epjc3]{svjour3}
\smartqed  % flush right qed marks, e.g. at end of proof

\RequirePackage{graphicx}
    \usepackage{amsmath,amssymb,enumerate,widetext}
    \usepackage{makeidx}
    \usepackage{amsfonts}
	
    \usepackage[usenames,dvipsnames]{pstricks}
    \usepackage{subfigure}
    \usepackage{epsfig}
    \usepackage{slashed}
\journalname{Eur. Phys. J. C}
\begin{document}

\title{Hidden Spin-3/2 Field in the Standard Model\thanksref{t1}
}


\author{Durmu\c{s} Demir\thanksref{e1,addr1}
        \and
        Canan  Karahan\thanksref{e2,addr1}
        \and
        Beste Korutlu\thanksref{e3,addr2}
        \and
        Ozan Sarg{\i}n\thanksref{e4,addr1}
}

\thankstext{t1}{This work is supported in part by the TUBITAK grant 115F212.}
%Grants or other notes
\thankstext{e1}{e-mail: demir@physics.iztech.edu.tr}
\thankstext{e2}{e-mail: cananduzturk@iyte.edu.tr}
\thankstext{e3}{e-mail: beste.korutlu@tubitak.gov.tr}
\thankstext{e4}{e-mail: ozansargin@iyte.edu.tr}

\institute{Department of Physics, \.{I}zmir Institute of Technology, Urla, \.{I}zmir, 35430 TURKEY \label{addr1}
           \and
           T\"{U}B\.{I}TAK National Metrology Institute, Gebze, Kocaeli, 41470 TURKEY \label{addr2}
           %\and
%           \emph{Present Address:} if needed\label{addr3}
}

\date{Received: \today / Accepted: \today}

\maketitle

\begin{abstract}
Here we show that a massive spin-3/2 field can hide in the
SM spectrum in a way revealing itself only virtually. We study 
collider signatures and loop effects of this field, 
and determine its role in Higgs inflation and 
its potential as Dark Matter. We show that this spin-3/2
field has a rich linear collider phenomenology and motivates
consideration of a neutrino-Higgs collider. We also show
that study of Higgs inflation, dark matter and dark energy 
can reveal more about the neutrino and dark sector.
\keywords{Spin-3/2 field, Linear Collider, Neutrino-Higgs Collider, Naturalness, Higgs inflation, Dark Matter}
\PACS{12.60.-i \and 95.35.+d}
\end{abstract}

\section{Introduction}
The Standard Model (SM) of strong and electroweak interactions, spectrally
completed by the discovery of its Higgs boson at the LHC \cite{higgs-mass}, 
seems to be the model of the physics at the Fermi energies. It does so because various experiments
have revealed so far no new particles beyond the SM spectrum. There is, however, at least the dark matter (DM), which requires new particles beyond the SM. Physically, therefore, we must use every 
opportunity to understand where those new particles can hide, if any.

In the present work we study a massive spin-3/2 field hidden in the SM spectrum. This higher-spin field, described by the Rarita-Schwinger equations \cite{Rarita:1941mf,pilling}, has to obey certain constraints to have correct degrees of freedom when it is on the physical shell. At the renormalizable level, it can couple to the SM matter via only the neutrino portal (the composite SM singlet formed by the lepton doublet and the Higgs field). This interaction is such that it vanishes when the spin-3/2 field is on shell. In Sec. 2 below we give the model and basic constraints on the spin-3/2 field. 

In Sec. 3 we study collider signatures of the spin-3/2 field. We study there $\nu_L h \rightarrow \nu_{L} h$ and $e^{-}e^{+}\rightarrow W^{+}W^{-}$  scatterings in detail. We give analytical computations and numerical predictions. We propose there a neutrino-Higgs collider and emphasize importance of the linear collider in probing the spin-3/2 field.

In Sec. 4 we turn to loop effects of the spin-3/2 field. We find that the spin-3/2 field adds logarithmic and quartic UV-sensitivities atop the logarithmic and quadratic ones in the SM. We convert  power-law UV-dependent terms into curvature terms as a result of the incorporation of gravity into the SM. Here we use the results of \cite{gravity,gravity2} which shows that gravity can be incorporated into the SM properly and naturally {\it (i)} if the requisite curved geometry is structured by interpreting the UV cutoff as a constant value assigned to the spacetime curvature, and {\it (ii)} if the SM is extended by a secluded new physics (NP) that does not have to interact with the SM. This mechanism eliminates big hierarchy problem by metamorphosing the quadratic UV part of the Higgs boson mass turns into Higgs-curvature coupling. 

In Sec. 5 we discuss possibility of Higgs inflation via the large Higgs non-minimal coupling induced by the spin-3/2 field. We find that Higgs inflation is possible in a wide range of parameters provided that the secluded NP sector is crowded enough. 

In Sec. 6 we discuss the DM. We show therein that the spin-3/2 field is a viable DM candidate. We also show that the singlet fields in the NP can form a non-interacting DM component. 

In Sec. 7 we conclude. There, we give a brief list of problems that can be studied as furthering of the material presented this work. 

\section{A Light Spin-3/2 Field}
\begin{figure}[h!]
  \begin{center}
  \includegraphics[scale=0.5]{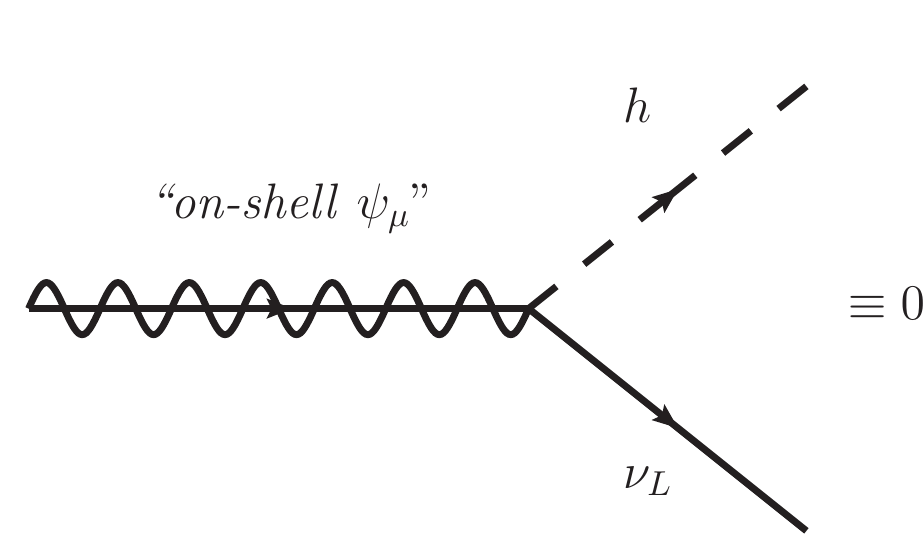}
  \end{center}
  \caption{$\psi_{\mu}-h-\nu_L$ coupling with vertex factor $i c_{3/2} \gamma^{\mu}$. Scatterings in which $\psi_{\mu}$ is on shell must all be forbidden since $c_{3/2} \gamma^{\mu} \psi_{\mu}$ vanishes on mass shell by the constraint (\ref{eqn4p}). This ensures stability of  $\psi_{\mu}$ against decays and all sort of co-annihilations.} \label{fig:Px}
\end{figure}

Introduced for the first time by
Rarita and Schwinger \cite{Rarita:1941mf}, $\psi_{\mu}$ propagates with
\begin{eqnarray}
S^{\alpha\beta}(p) = \frac{i}{{\slashed{p}} - M} \Pi^{\alpha\beta}(p),
\end{eqnarray}
to carry one spin-3/2 and two spin-1/2 components through the
projector \cite{pilling}
\begin{eqnarray}
\label{project}
\Pi^{\alpha\beta} = -\eta^{\alpha\beta} +
\frac{\gamma^{\alpha}\gamma^{\beta}}{3}+
\frac{\left(\gamma^{\alpha}p^{\beta} -
\gamma^{\beta}p^{\alpha}\right)}{3M}+\frac{2
p^{\alpha}p^{\beta}}{3 M^2},
\end{eqnarray}
that exhibits both spinor and vector characteristics.
It is necessary to impose \cite{pilling}
\begin{eqnarray}
\label{eqn4}
p^{\mu}\psi_{\mu}(p)\rfloor_{p^2=M^2}=0,
\end{eqnarray}
and
\begin{eqnarray}
\label{eqn4p}
\gamma^{\mu}\psi_{\mu}(p)\rfloor_{p^2=M^2}=0,
\end{eqnarray}
to eliminate the two spin-1/2 components to make $\psi_{\mu}$
satisfy the Dirac equation
\begin{eqnarray}\label{eqn5}
\left(\slashed{p} - M\right)\psi_{\mu}=0
\end{eqnarray}
as expected of an on-shell fermion. The constraints (\ref{eqn4}) and (\ref{eqn4p}) imply that $p^{\mu}\psi_{\mu}(p)$ and $\gamma^{\mu}\psi_{\mu}(p)$ both vanish on the physical shell $p^2=M^2$. The latter is illustrated in Fig. \ref{fig:Px} taking $\psi_{\mu}$ on-shell.

Characteristic of singlet fermions, the $\psi_{\mu}$, at the renormalizable level, makes contact with the SM via 
\begin{eqnarray}
\label{int1}
{\mathcal{L}}^{(int)}_{3/2} = c^{i}_{{3/2}} \overline{L^{i}} H \gamma^{\mu}\psi_{\mu} + {\text{h.c.}}
\end{eqnarray}
in which 
\begin{eqnarray}
L^i = \left(\begin{array}{c}\nu_{\ell L}\\ \ell_L\end{array}\right)_{i}
\end{eqnarray}
is the lepton doublet ($i=1,2,3$), and 
\begin{eqnarray}
H = \frac{1}{\sqrt{2}}\left(\begin{array}{c}v + h + i \varphi^0\\ \sqrt{2} \varphi^{-}\end{array}\right) 
\end{eqnarray}
is the Higgs doublet with vacuum expectation value $v\approx 246\ {\rm GeV}$, Higgs boson $h$, and Goldstone bosons  $\varphi^{-}$, $\varphi^0$ and $\varphi^+$ (forming the longitudinal components of $W^{-}$, $Z$ and $W^{+}$ bosons, respectively). 

In general, neutrinos are sensitive probes of singlet fermions. They can get masses through, for instance, the Yukawa interaction (\ref{int1}), which leads to the Majorana mass matrix 
\begin{eqnarray}
(m_{\nu})^{i j}_{3/2} \propto c^i_{{3/2}} \frac{v^2}{M} c^{\star j}_{{3/2}} 
\end{eqnarray}
after integrating out $\psi_{\mu}$. This mass matrix can, however, not lead to the experimentally known neutrino mixings \cite{neutrino-mass}. This means that flavor structures necessitate additional singlet fermions. Of such are the right-handed neutrinos  $\nu_R^k$ of mass $M_k$ ($k=1,2,3,\dots$), which interact with the SM through 
\begin{eqnarray}
\label{int2}
{\mathcal{L}}^{(int)}_{R} = c_{{R}}^{i k} \bar{L}^i H \nu_R^k + {\text{h.c.}}
\end{eqnarray}
to generate the neutrino Majorana masses 
\begin{eqnarray}
(m_{\nu})^{i j}_{R} \propto c_{{R}}^{i k} \frac{v^2}{M_k} c_{{R}}^{\star k j}
\end{eqnarray}
of more general flavor structure. This mass matrix must have enough degrees of freedom  to fit to the data \cite{neutrino-mass}.

\begin{figure}[h!]
  \begin{center}
  \includegraphics[scale=0.45]{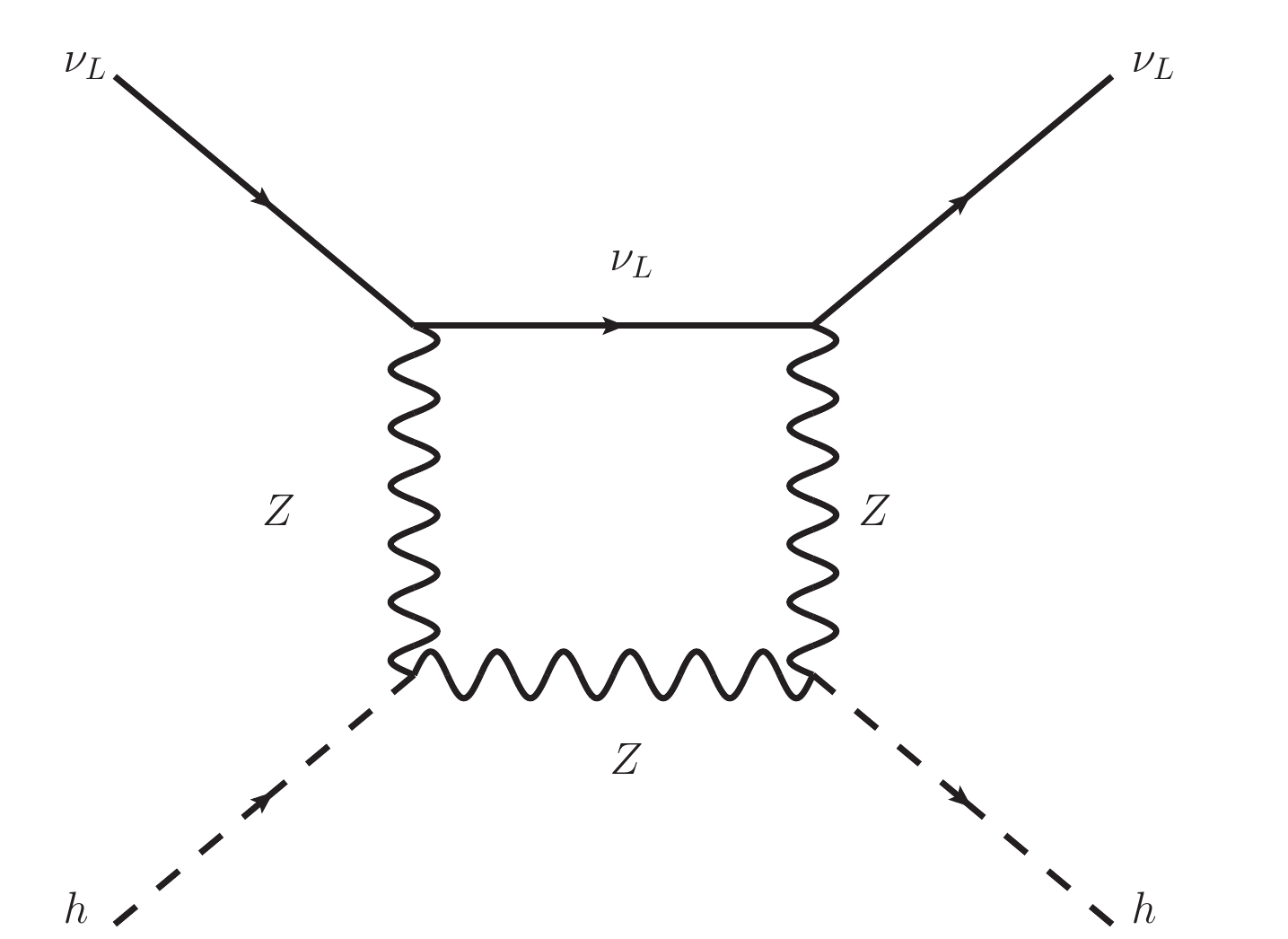}
%  \end{center}
%  \caption{The Z-mediated box diagram for $\nu h \rightarrow \nu h$.} \label{fig:zx}
%\end{figure}
%
%
%\begin{figure}[h!]
%  \begin{center}
%  \includegraphics[scale=0.45]{vhvhWmed-cropped.pdf}
  \end{center}
  \caption{The $\nu-Z$ box mediating the $\nu_L h \rightarrow \nu_L h$ scattering in the SM. The $e-W$ box is not shown. } \label{nhnh-SM}
\end{figure}

\begin{figure}[ht!]
  \begin{center}
  \includegraphics[scale=0.40]{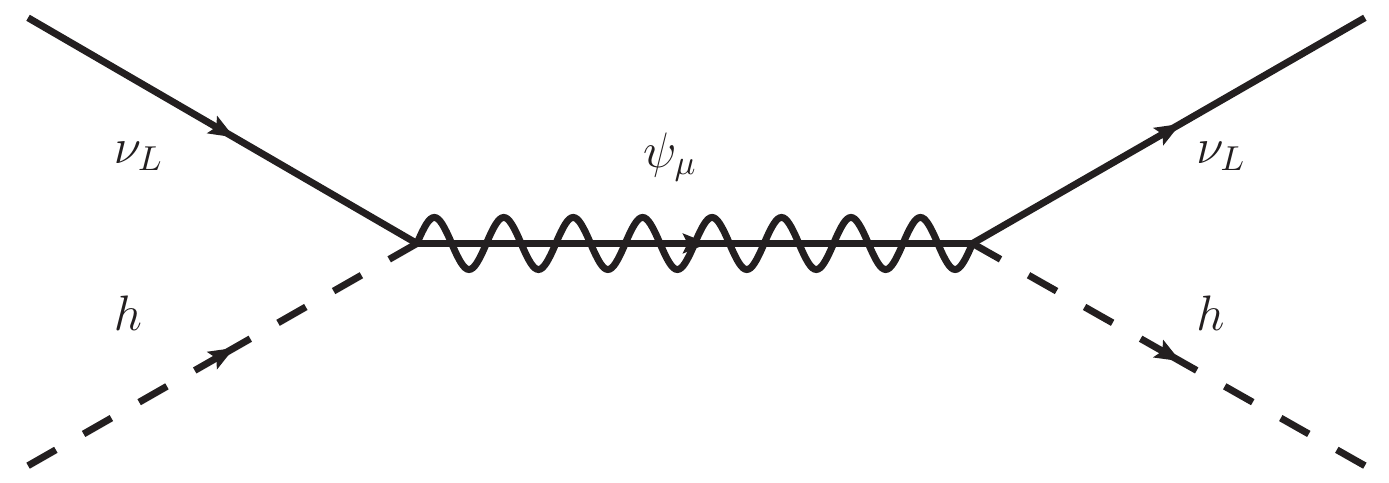}
  \end{center}
  %  \vspace{-1cm}
  \caption{$\nu_L h \rightarrow \nu_L h$ scattering with $\psi_{\mu}$ mediation. No resonance can occur  at $\sqrt{s}=M$ because $\psi_{DM}$ cannot come to mass shell.} \label{nhnh-3/2}
\end{figure}
Here we make a pivotal assumption. We assume that $\psi_{\mu}$ and $\nu_R^k$ can weigh as low as a TeV, and that $c^i_{{3/2}}$ and some of $c_{{R}}^{i k}$ can be ${\mathcal{O}}(1)$. We, however, require that contributions to neutrino masses from 
$\psi_{\mu}$ and $\nu_R$ add up to reproduce with experimental result
\begin{eqnarray}
\label{numass}
(m_{\nu})^{i j}_{3/2} + (m_{\nu})^{i j}_{R} \approx (m_{\nu})^{i j}_{exp} 
\end{eqnarray}
via cancellations among different terms.  We therefore take 
\begin{eqnarray}
c_{{3/2}} \lesssim {\mathcal{O}}(1)\,,\; M\gtrsim {\rm TeV}
\end{eqnarray}
and investigate the physics of  $\psi_{\mu}$. This cancellation requirement does not have to cause any excessive fine-tuning simply because $\psi_{\mu}$ and $\nu_R^k$  can have appropriate symmetries that correlate their couplings. One possible symmetry would be rotation of $\gamma^{\mu}\psi_{\mu}$ and $\nu_R^k$ into each other. We defer study of possible symmetries to another work in progress \cite{Ozan}. The right-handed sector, which can involve many $\nu_R^k$ fields, is interesting by itself but hereon we focus on $\psi_{\mu}$ and take, for simplicity, $c^i_{{3/2}}$ real and family-universal ($c^i_{{3/2}}=c_{{3/2}}$ for $\forall$ $i$).

\section{Spin-3/2 Field at Colliders}
It is only when it is off-shell that $\psi_{\mu}$ can reveal itself through the interaction (\ref{int1}). This means that its effects are restricted to modifications in scattering rates of the SM particles. To this end, as follows from (\ref{int1}), it participates in 
\begin{enumerate}
\item $\nu_L h \rightarrow \nu_{L} h$ (and also $\nu_{L}\nu_{L} \rightarrow h h$)
\item $e^+ e^- \rightarrow W^+_L W^-_L$ (and also $\nu_{L}\nu_{L} \rightarrow Z_L Z_L$)
\end{enumerate}
at the tree level. They are analyzed below in detail.

\subsection{$\nu_L h \rightarrow \nu_{L} h$ Scattering}
Shown in Fig. \ref{nhnh-SM} are the two box diagrams which enable $\nu_L h \rightarrow \nu_L h$ scattering in the SM.  Added to this loop-suppressed SM piece is the $\psi_{\mu}$ piece depicted in Fig. \ref{nhnh-3/2}. The two contributions add up to give the cross section 
\begin{eqnarray}
\frac{d\sigma(\nu_L h \rightarrow \nu_L h)}{dt}= \frac{1}{16\pi}\frac{{\mathcal{T}_{\nu h}}({{s}},{{t}})}{(s-m_{h}^2)^2}
\end{eqnarray}
in which the squared matrix element 

\begin{widetext}
\begin{eqnarray}
\label{mat-el-nuhnuh}
{\mathcal{T}_{\nu h}}({{s}},{{t}}) &=& 9\! \left(\frac{c_{3/2}}{3 M}\right)^4\!\! \left(\!
\left({{s}}-m_h^2\right)^2 + {{s}}{{t}}\right) \!-\! 16\! \left(\frac{c_{3/2}}{3 M}\right)^2\!\! \left(\!
2\left({{s}}-m_h^2\right)^2 \!+\! \left(2{{s}} -m_h^2\right){{t}}\right) {\mathbb{L}} \!+\! 2\left(
{{s}}-m_h^2\right)\left({{s}} + {{t}}-m_h^2\right) {\mathbb{L}}^2
\end{eqnarray}
\end{widetext}
\noindent involves the loop factor
\begin{eqnarray}
{\mathbb{L}}=\! \frac{(g_W^2\!+\!g_Y^2)^2 M_Z^2 m_h^2 I(M_Z)}{192 \pi^2}\! + \!\frac{g_W^4 M_W^2 m_h^2 I(M_W)}{96 \pi^2}
\end{eqnarray}
in which $g_W$ ($g_Y$) is the isospin (hypercharge) gauge coupling, and

\begin{figure}[h!]
  \begin{center}
  \includegraphics[scale=1.5]{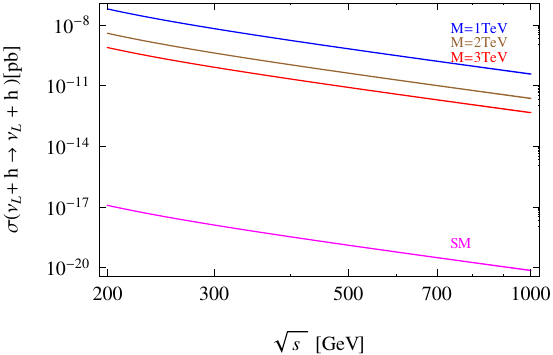}
  \end{center}
  \caption{The total cross section for $\nu_L h \rightarrow \nu_L h$ scattering as a function of the neutrino-Higgs center-of-mass energy $\sqrt{s}$ for $M=1, 2$ and $3\ {\rm TeV}$ at $c_{3/2}= 1$. Cases with  $c_{3/2}\neq 1$ can be reached via the rescaling $M\rightarrow M/c_{3/2}$.} \label{fig:Pxx}
\end{figure}

\begin{widetext}
\begin{eqnarray}
I(\mu)=\int_{0}^{1}dx\int_{0}^{1-x}dy\int_{0}^{1-x-y}dz \left((s-m_h^2)(x+y+z-1) y - txz + m_h^2 y (y-1) + \mu^2 (x + y + z)\right)^{-2}
\end{eqnarray}
\end{widetext}
\noindent is the box function. In Fig. \ref{fig:Pxx}, we plot the total cross section $\sigma(\nu_L h \rightarrow \nu_L h)$  as a function of the neutrino-Higgs center-of-mass energy for different $M$ values. The first important thing about the plot is that there is no resonance formation around $\sqrt{s}=M$. This confirms the fact that $\psi_{\mu}$, under the constraint (\ref{eqn4p}), cannot come to physical shell with the couplings in (\ref{int1}). In consequence, the main search strategy for $\psi_{\mu}$ is to look for deviations from the SM rates rather than resonance shapes. The second important thing about the plot is that, in general, as revealed by (\ref{mat-el-nuhnuh}), larger the $M$ smaller the $\psi_{\mu}$ contribution. The cross section starts around $10^{-7}\ {\rm pb}$, and falls rapidly with $\sqrt{s}$. (The SM piece, as a loop effect, is too tiny to be observable: $\sigma(\nu_L h \rightarrow \nu_L h)\lesssim 10^{-17}\ {\rm pb}$). It is necessary to have some $10^{4}/fb$ integrated luminosity (100 times the target luminosity at the LHC) to observe few events in a year. This means that $\nu_L \nu_L \rightarrow h h$ scattering can probe $\psi_{\mu}$ at only high luminosity but with a completely new scattering scheme. 
 
\begin{figure}[ht!]
  \begin{center}
  \includegraphics[scale=0.50]{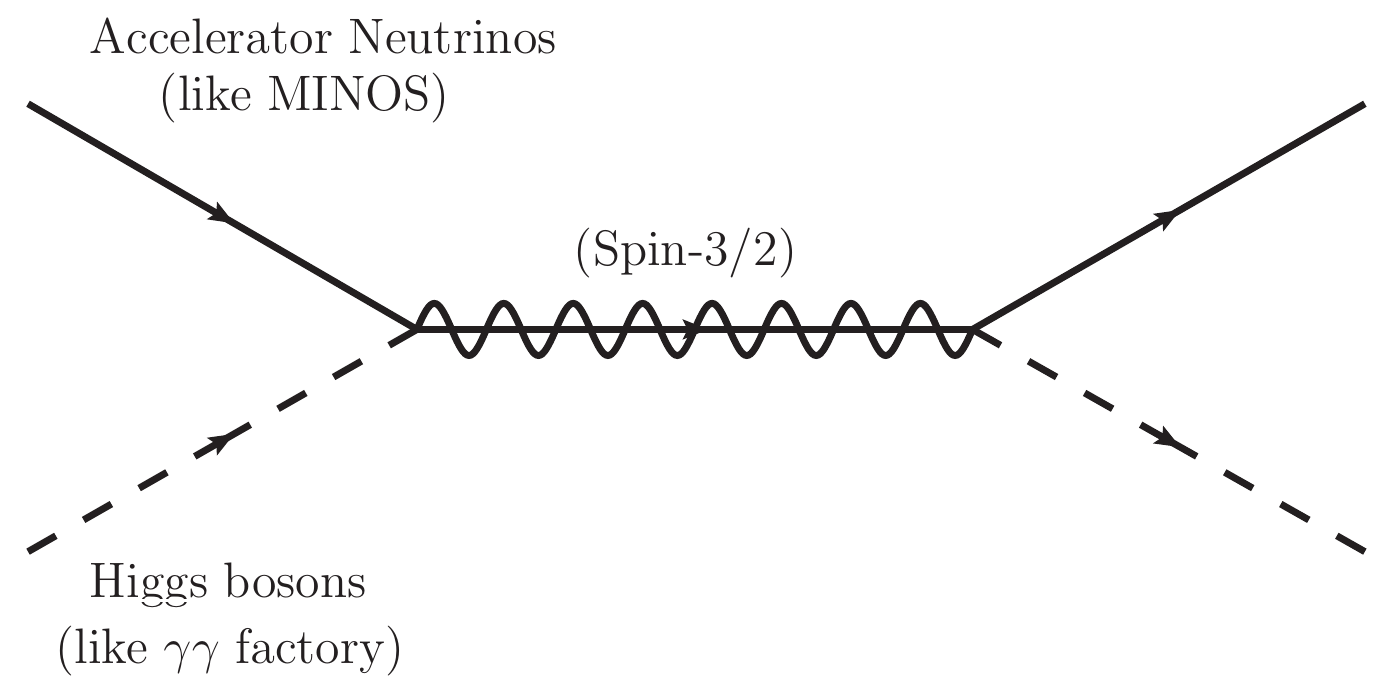}
  \end{center}
  \caption{Possible neutrino-Higgs collider to probe $\psi_{\mu}$.} \label{fig:P10}
\end{figure}
Fig. \ref{fig:Pxx} shows that neutrino-Higgs scattering can be a promising channel to probe $\psi_{\mu}$ (at high-luminosity, high-energy machines). The requisite experimental setup would involve crossing of Higgs factories with accelerator neutrinos. The setup, schematically depicted in  Fig. \ref{fig:P10}, can be viewed as incorporating  future Higgs (CEPC\cite{Ruan:2014xxa}, FCC-ee \cite{Gomez-Ceballos:2013zzn} and ILC \cite{Baer:2013cma}) and neutrino \cite{Choubey:2011zzq} factories. If ever realized, it could be a rather clean experiment with negligible SM background. This hypothetical ``neutrino-Higgs collider'', depicted in Fig. \ref{fig:P10}, must have, as suggested by Fig. \ref{fig:Pxx}, some $10^4/fb$ integrated luminosity to be able to probe a TeV-scale $\psi_{\mu}$. In general, need to high luminosities is a disadvantage of this channel.  (Feasibility study, technical design and possible realization of a ``neutrino-Higgs collider'' falls outside the scope of the present work.)

\begin{figure}[ht!]
 \begin{center}
  \includegraphics[scale=0.45]{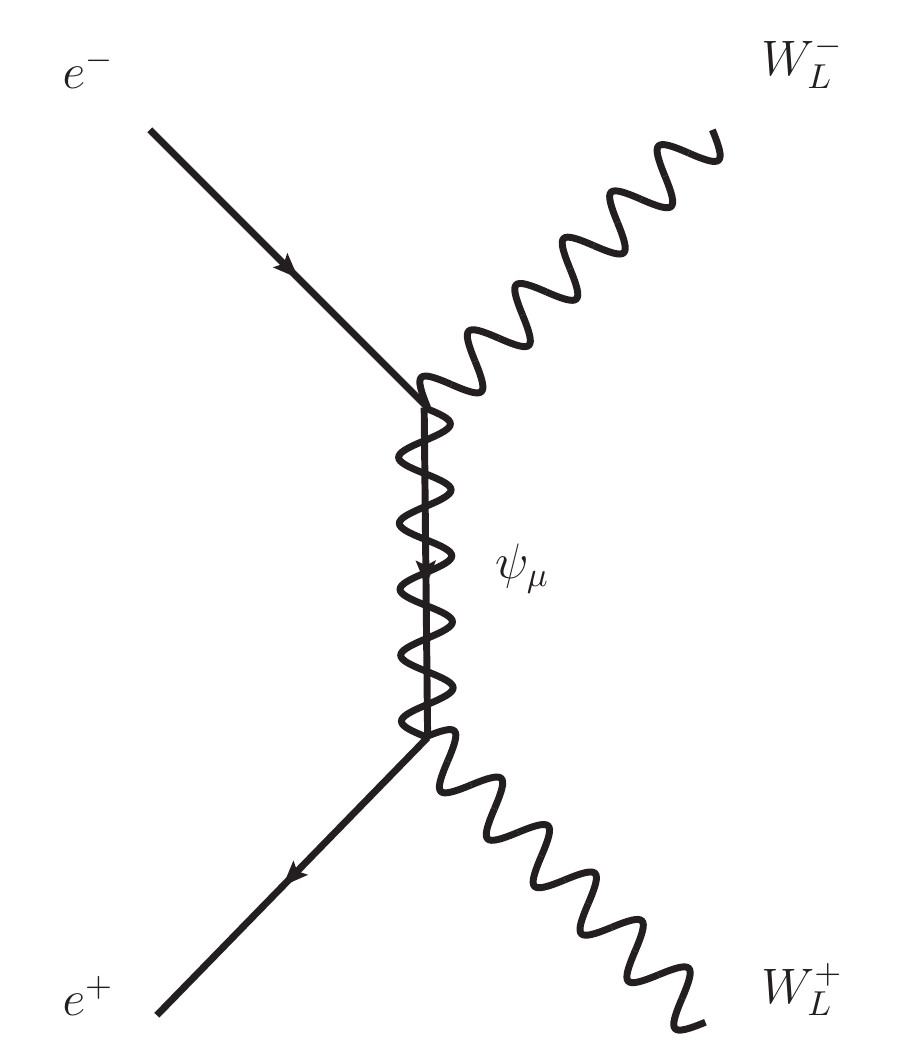}
 \end{center}
  \caption{The Feynman diagram for $e^+ e^- \rightarrow W_L^+ W_L^-$ scattering. The $\nu_L \nu_L \rightarrow Z_L Z_L$ scattering has the same topology.} \label{fig:w6}
\end{figure}

\subsection{$e^+ e^- \rightarrow W_L^+ W_L^-$ Scattering}
It is clear that $\psi_{\mu}$ directly couples to the Goldstone bosons $\varphi^{+,-,0}$ via (\ref{int1}). The Goldstones, though eaten up by the $W$ and $Z$ bosons in acquiring their masses, reveal themselves at high energies. In fact, the Goldstone equivalence theorem \cite{equivalence} states that scatterings at energy $E$ involving longitudinal $W^{\pm}_L$ bosons are equal to scatterings that involve $\varphi^{\pm}$ up to terms ${\mathcal{O}}(M_W^2/E^2)$. This theorem, with similar equivalence for the longitudinal $Z$ boson, provides a different way of probing $\psi_{\mu}$. In this regard, depicted in  Fig. \ref{fig:w6} is $\psi_{\mu}$ contribution to
$e^+ e^- \rightarrow W_L^+ W_L^-$ scattering in light of the Goldstone equivalence. The SM amplitude is  given in \cite{equivalence}. The total differential cross section
\begin{eqnarray}
\frac{d\sigma(e^+ e^- \rightarrow W^+_L W^-_L)}{dt}= \frac{1}{16\pi s^2} {{\mathcal{T}_{W_L W_L}}({{s}},{{t}})}
\end{eqnarray}
involves the squared matrix element 

\begin{figure}[h!]
  \begin{center}
  \includegraphics[scale=1.45]{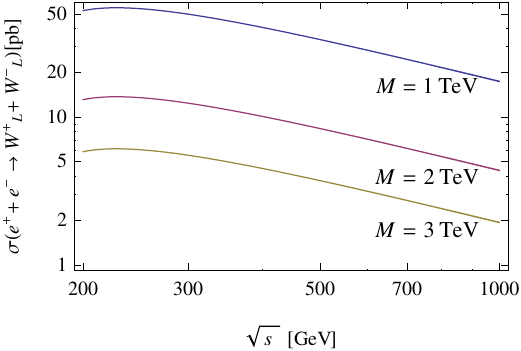}
  \end{center}
  \caption{ The total cross section for $e^{-}e^{+}\rightarrow W^{+}W^{-}$ scattering as a function of the electron-positron center-of-mass energy $\sqrt{s}$ for $M=1, 2$ and $3\ {\rm TeV}$ at $c_{3/2}= 1$. Cases with  $c_{3/2}\neq 1$ can be reached via the rescaling $M\rightarrow M/c_{3/2}$.} \label{fig:Wxx}
\end{figure}

\begin{widetext}
\begin{eqnarray}
\label{mat-el-nuhnuh}
{{\mathcal{T}_{W_L W_L}}({{s}},{{t}})}\! &=&\! \left(\! \frac{g_W^2}{s-M_Z^2}\left(\!-1+\! \frac{M_Z^2}{4 M_W^2}\! +\! \frac{M_Z^2-M_W^2}{s}\right)\! +\!  \frac{g_W^2}{s-4 M_Z^2}\left(\!1+\! \frac{M_W^2}{t}\right)\! +\! \frac{c^{2}_{3/2}}{3 M^2}\right)^{2}\!\!\! \left(-2 s M_W^2 -2 (t-M_W^2)^2\right) \nonumber\\
&+&\frac{c^4_{3/2} s}{18 M^2} \left(4 + \frac{t}{t-M^2}\right)^2
\end{eqnarray}
\end{widetext}
\noindent Plotted in Fig. \ref{fig:Wxx} is $\sigma(e^+ e^- \rightarrow W^+_L W^-_L)$ as a function of the $e^+ e^-$ center-of-mass energy for different values of $M$. The cross section, which falls with $\sqrt{s}$ without exhibiting a resonance shape, is seen to be large enough to be measurable at the ILC \cite{Baer:2013cma}. In general, larger the $M$ smaller the cross section but even $1/fb$ luminosity is sufficient for probing $\psi_{\mu}$ for a wide range of mass values.
Collider searches for $\psi_{\mu}$, as illustrated by  $\nu_L h \rightarrow \nu_{L} h$ and $e^{-}e^{+}\rightarrow W^{+}W^{-}$ scatterings, can access spin-3/2 fields of several TeV mass. For instance, the ILC, depending on its precision, can confirm or exclude a $\psi_{\mu}$ of even 5 TeV mass with an integrated luminosity around $1/fb$. Depending on possibility and feasibility of a neutrino-neutrino collider (mainly accelerator neutrinos), it may be possible to study also $\nu_L \nu_L \rightarrow h h$ and $\nu_L \nu_L \rightarrow Z_L Z_L$ scatterings, which are expected to have similar sensitivities to $M$. 

\section{Spin-3/2 Field in Loops}
As an inherently off-shell field, $\psi_{\mu}$ is expected to reveal itself mainly in loops. Its one possible loop effect would be generation of neutrino masses but chirality forbids it. Despite the couplings in (\ref{int1}), therefore, neutrino masses do not get any contribution from $\psi_{\mu}-h$ loop.

\begin{figure}[ht!]
  \begin{center}
  \includegraphics[scale=0.45]{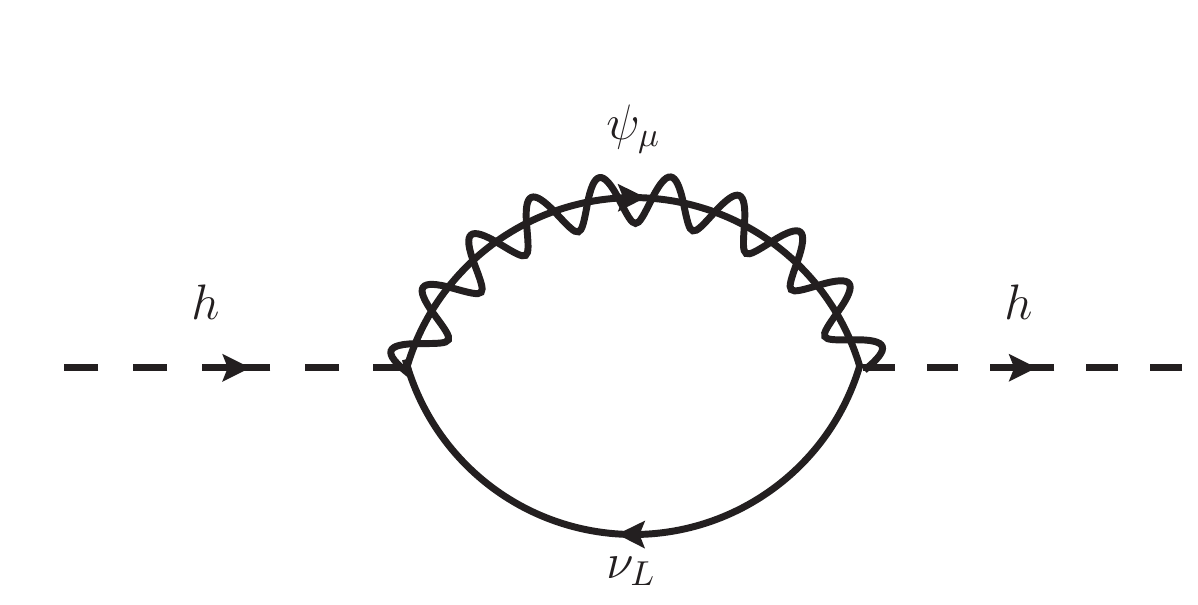}
  \end{center}
  %  \vspace{-1cm}
  \caption{The $\psi_{\mu}-\nu_L$ loop that generates the logarithmic correction in (\ref{eqn8}) and the quartic correction in (\ref{eqn88}).} \label{fig:P7}
\end{figure}

One other loop effect of $\psi_{\mu}$ would be radiative corrections to the Higgs boson mass. This is not forbidden by any symmetry. The relevant Feynman diagram is depicted in Fig. \ref{fig:P7}. It adds to the Higgs boson squared-mass a logarithmic piece
\begin{eqnarray}
\label{log-corr}
\left(\delta m_h^2\right)_{log} = \frac{c_{3/2}^2}{12\pi^2}M^2\log G_F M^2
\end{eqnarray}
relative to the logarithmic piece $\log G_F \Lambda^2$ in the SM, and a quartic piece
\begin{eqnarray}\label{eqn88}
\left(\delta m_h^2\right)_{4} = \frac{c_{3/2}^2}{ 48 \pi^2} \frac{ \Lambda^4}{M^2}
\end{eqnarray}
which have the potential to override the experimental result \cite{higgs-mass} depending on how large the UV cutoff $\Lambda$ is compared to the Fermi scale $G_F^{-1/2} = 293\ {\rm GeV}$. 

The logarithmic contribution in (\ref{log-corr}), which originates from the $\eta^{\alpha\beta}$ part of (\ref{project}), gives rise to the little hierarchy problem in that larger the $M$ stronger the destabilization of the SM Higgs sector. Leaving aside the possibility of cancellations with similar contributions from the right-handed neutrinos $\nu_R^k$ in (\ref{int2}), the little hierarchy problem can be prevented if $M$ (more precisely $M/c_{3/2}$) lies in the TeV domain. 

The quartic contribution in (\ref{eqn88}), which originates from the longitudinal $p^{\alpha} p^{\beta}$ term in (\ref{project}), gives cause to the notorious big hierarchy problem in that larger the $\Lambda$ larger the destabilization of the SM Higgs sector. This power-law UV sensitivity exists already in the SM
\begin{eqnarray}\label{eqn8}
\left(\delta m_h^2\right)_{2}&=&\frac{3 \Lambda^2}{16 \pi^2 {\left|\langle H \rangle\right|^2}}\left( m_h^2 + 2 M_W^2 + M_Z^2 - 4 m_t^2\right)
\end{eqnarray}
at the quadratic level \cite{Veltman:1980mj} and
violates the LHC bounds unless $\Lambda \lesssim 550\
{\rm GeV}$. This bound obviously contradicts with the
LHC experiments since the latter continue to confirm the
SM at multi {\rm TeV} energies. This experimental fact makes
it obligatory to find a natural UV completion to the SM.

One possibility is to require  $\left(\delta m_h^2\right)_{4}$ to cancel out  $\left(\delta m_h^2\right)_{2}$. This requirement involves a severe fine-tuning (as with a scalar field
\cite{fine-tune-scalar}, Stueckelberg vector \cite{besteyle} and
spacetime curvature \cite{curvature-ft}) and cannot form a viable
stabilization mechanism.

Another possibility would be to switch, for instance, to dimensional
regularization scheme, wherein the quartic and quadratic
UV-dependencies are known to disappear. This, however, is not a
solution. The reason is that the SM, as a quantum field theory of
the strong and electroweak interactions, needs gravity to be
incorporated as the forth known force. And the fundamental scale of
gravity, $M_{Pl}$, inevitably sets an ineliminable physical UV
cutoff (rendering $\Lambda$ physical). This cutoff forces quantum field theories to exist in between
physical UV and IR scales. The SM plus $\psi_{\mu}$ (plus right-handed neutrinos), for instance,
ranges from $G_{F}^{-1/2}$ at the IR up to $\Lambda$ at the UV such
that both scales are physical (not to be confused with the formal
momentum cutoffs employed in the cutoff regularization).

To stabilize the SM, it is necessary to metamorphose the
destabilizing UV effects. This necessitates a physical agent. The
most obvious candidate is gravity. That is to say, the
UV-naturalness problems can be a clue to how quantized matter must
gravitate. Indeed, quantized matter in classical curved geometry
suffers from inconsistencies. The situation can be improved by
considering long-wavelength matter by integrating out high-frequency
modes. This means that the theory to be carried into curved geometry
for incorporating gravity is not the full action but the effective
action (see the discussions in \cite{gravity} and \cite{gravity2}). Thus, starting with
the SM effective action in flat spacetime with well-known
logarithmic, quartic and quadratic UV-sensitivities, gravity can be
incorporated in a way ensuring UV-naturalness. More precisely,
gravity gets incorporated properly and naturally  {\it (i)} if the
requisite curved geometry is structured by interpreting $\Lambda^2$
as a constant value assigned to the spacetime curvature, and {\it
(ii)} if the SM is extended by new physics (NP) that does not have
to interact with the SM. The $\psi_{\mu}$ can well be an NP field.  
Incorporating gravity by identifying $\Lambda^2 g_{\mu\nu}$ with
the Ricci curvature $R_{\mu\nu}(g)$, fundamental scale of 
gravity gets generated as
\begin{eqnarray}
\label{MPl}
M_{Pl}^2 \approx \frac{\left(n_b-n_f\right)}{2(8 \pi)^2} \Lambda^2
\end{eqnarray}
where $n_b$ ($n_f$) are the total number of bosons (fermions) in the
SM plus the NP. The $\psi_{\mu}$ increases $n_f$ by 4, right-handed neutrinos by 2. There are
various other fields in the NP, which contribute to $n_b$  and $n_f$
to ensure $\Lambda \lesssim M_{Pl}$. Excepting $\psi_{\mu}$,  they
do not need to interact with the SM fields. Induction of $M_{Pl}$
ensures that the quadratic UV-contributions to vacuum energy are
canalized not to the cosmological constant but to the gravitational
constant (see  \cite{demir-ccp} arriving at this result in a
different context). This suppresses the cosmological constant down
to the neutrino mass scale.

The quartic UV contributions in (\ref{eqn88}) and the quadrat\-ic
contributions in (\ref{eqn8}) (suppressing contributions from the right-handed 
neutrinos $\nu_R^k$) change their roles with the inclusion
of gravity. Indeed, corrections to the Higgs mass term $\left[\left(\delta m_h^2\right)_{4}\!+\!\left(\delta m_h^2\right)_{2} \right]\!\! H^{\dagger}\! H$  turns
into
\begin{equation}
\label{exp}
\left[\!\frac{3\!\left(\!m_h^2\! +\! 2 M_W^2\! +\! M_Z^2\! -\! 4 m_t^2\!\right)}{(8\pi)^2\left|\langle H \rangle\right|^2}
\!+\! \frac{c_{3/2}^2}{12(n_b\!-\!n_f)}\! \frac{M_{Pl}^2}{M^2}\! \right]\!\! R H^{\dagger}\! H
\end{equation}
which is nothing but the direct coupling of the Higgs field to the
scalar curvature $R$. This Higgs-curvature coupling is perfectly
natural; it has no potential to de-stabilize the Higgs sector.
Incorporation of gravity as in \cite{gravity,gravity2} leads, therefore, to
UV-naturalization of the SM with a nontrivial NP sector
containing $\psi_{\mu}$ as its interacting member. 

\section{Spin-3/2 Field as Enabler of Higgs Inflation}
The non-minimal Higgs-curvature coupling in (\ref{exp}) reminds one at once the possibility of Higgs inlation. Indeed, the Higgs field has been shown in \cite{higgs-inf,higgs-inf-2} to lead to correct inflationary expansion provided that 
\begin{eqnarray}
\frac{c_{3/2}^2}{12(n_b-n_f)} \frac{M_{Pl}^2}{M^2} \approx 1.7\times 10^{4}
\end{eqnarray}
after dropping the small SM contribution in (\ref{exp}). This relation puts constraints on $M$ and $\Lambda$ depending on how crowded the NP is. 

For a Planckian UV cutoff $\Lambda \approx M_{Pl}$, the Planck scale in (\ref{MPl}) requires $n_b - n_f\approx 1300$, and this leads to $M/c_{3/2}\approx 6.3\times 10^{13}\ {\rm GeV}$. This heavy $\psi_{\mu}$, weighing not far from the see-saw and axion scales, acts as an enabler of Higgs inflation. (Of course, all this makes sense if the $\psi_{\mu}$ contribution in (\ref{log-corr}) is neutralized by similar contributions from the right-handed neutrinos $\nu_R^k$ to alleviate the little 
hierarchy problem.) 

For an intermediate UV cutoff $\Lambda\ll M_{Pl}$, $n_b-n_f$ can be large enough to bring $M$ down to lower scales. In fact, $M$ gets lowered to $M\sim {\rm TeV}$ for $n_b-n_f\simeq 10^{24}$, and this sets the UV cutoff $\Lambda \sim 3\ {\rm TeV}$. This highly crowded NP illustrates how small $M$
and $\Lambda$ can be. Less crowded NP sectors lead to intermediate-scale $M$ and $\Lambda$.

It follows therefore that it is possible to realize Higgs inflation through the Higgs-curvature coupling (corresponding to quartic UV-dependence the $\psi_{\mu}$ induces on the Higgs mass). It turns out that Higgs inflation is decided by how heavy $\psi_{\mu}$ is and how crowded the NP is. It is interesting that the $\psi_{\mu}$ hidden in the SM spectrum enables successful Higgs inflation if gravity is incorporated into the SM as in \cite{gravity,gravity2}. 

\section{Spin-3/2 Field as Dark Matter} 
Dark matter (DM), forming one-forth of the matter in the Universe, must be electrically-neutral and long-lived. The negative searches \cite{plehn,leszek} so far have added one more feature: The DM must have exceedingly suppressed interactions with the SM matter. It is not hard to see that the spin-3/2 fermion $\psi_{\mu}$ possesses all these properties.  Indeed, the constraint (\ref{eqn4p}) ensures that scattering processes in which $\psi_{\mu}$ is on its mass shell must all be forbidden simply because its interactions in (\ref{int1})  involves the vertex factor $c_{3/2} \gamma^{\mu}$. This means that decays of $\psi_{\mu}$ as in Fig.\ref{fig:Px} as well as its co-annihilations with the self and other SM fields are all forbidden. Its density therefore does not change with time, and the observed DM relic density \cite{planck} must be its primordial density, which is determined by the  short-distance physics the $\psi_{\mu}$ descends from. It is not possible to calculate the relic density without knowing the short-distance physics. Its mass and couplings, on the other hand, can be probed via the known SM-scatterings as studied in Sec. 3 above. In consequence, the $\psi_{\mu}$, as an inherently off-shell fermion hidden in the SM spectrum, possesses all the features required of a DM candidate.  

Of course, the $\psi_{\mu}$ is not the only DM candidate in the setup. The crowded NP sector, needed to incorporate gravity in a way solving the hierarchy problem (see Sec. 4 above), involves various fields which do not interact with the SM matter. They are viable candidates for non-ineracting DM as well as dark energy (see the detailed analysis in \cite{gravity2}). The non-interacting NP fields can therefore contribute to the total DM distribution in the Universe. It will be, of course, not possible to search for them directly or indirectly. In fact, they do not have to come to equilibrium with the SM matter. 

Interestingly, both $\psi_{\mu}$ and the secluded fields in the NP act as extra fields hidden in the SM spectrum. Unlike $\psi_{\mu}$ which reveal itself virtually, the NP singlets remain completely intact. The main implication is that, in DM phenomenology, one must keep in mind that there can exist an unobservable, undetectable component of the DM \cite{gravity2}.

\section{Conclusion and Outlook}
In this work we have studied a massive spin-3/2 particle $\psi_{\mu}$ obeying the constraint (\ref{eqn4p}) and interacting with the SM via (\ref{int1}). It hides in the SM spectrum as an
inherently off-shell field. We first discussed its collider signatures by studying $\nu_L h \rightarrow \nu_{L} h$ and $e^{-}e^{+}\rightarrow W^{+}W^{-}$ scatterings in detail in Sec. 3. Following this, we turned to its loop effects and determined how it contributes to big and little hierarchy problems in the SM. Resolving the former by appropriately incorporating gravity, we show that Higgs field can inflate the Universe. Finally, we show that $\psi_{\mu}$ is a viable 
DM candidate, which can be indirectly probed via the scattering processes we have analyzed. 

The material presented in this work can be extended in various ways. A partial list would include:
\begin{itemize}
\item Determining under what conditions right-handed neutrinos can lift the constraints on $\psi_{\mu}$ from the neutrino masses,

\item Improving the analyses of $\nu_L h \rightarrow \nu_{L} h$ and $e^{-}e^{+}\rightarrow W^{+}W^{-}$ scatterings by including loop contributions,

\item Simulating $e^{-}e^{+}\rightarrow W^{+}W^{-}$ at the ILC by taking into account planned detector acceptances and collider energies,

\item Performing a feasibility study of the proposed neutrino-Higgs collider associated with $\nu_L h \rightarrow \nu_{L} h$ scattering,

\item Exploring UV-naturalness by including right-handed neutrinos, and determining under what conditions the little hierarchy problem is softened.

\item Including effects of the right-handed neutrinos into Higgs inflation, and determining appropriate parameter space. 

\item Giving an in-depth analysis of the dark matter and dark energy by taking into account the spin-3/2 field, right-handed neutrinos and the secluded NP fields.  

\item Studying constraints on the masses of NP fields from nucleosynthesis and other processes in the early Universe. 
\end{itemize}
We will continue to study the spin-3/2 hidden field starting with some of these points.

{\bf Acknowledgements.}
This work is supported in part by the TUBITAK grant 115F212. We thank to the conscientious referee for enlightening comments and suggestions.

\end{document}